• Article •

# Evidence for rotational to vibrational evolution along the yrast line in the odd-A rare-earth nuclei

H. B. Zhou[1*], S. Huang[1], G. X. Dong[2, 3], Z. X. Shen[1], H. J. Lu[1], L. L. Wang[1], X. J. Sun[1] and F. R. Xu[3]

[1]*Guangxi Normal university, Guilin 541004, People's Republic of China*
[2]*School of Science, Huzhou University, Huzhou 313000, China*
[3]*School of Physics and SK Laboratory of Nuclear Physics & Technology, Peking University, Beijing 100871, People's Republic of China*

The phase transition of nuclei to increasing angular momentum (or spin) and excitation energy is one of the most fundamental topics of nuclear structure research. The odd-$N$ nuclei with $A \approx 160$ are widely considered belonging to the well-deformed region, and their excitation spectra are energetically favored to exhibit the rotational characteristics. In the present work, however, there is evidence indicating that the nuclei can evolve from rotation to vibration along the yrast lines while increasing spin. The simple method, named as $E$-Gamma Over Spin ($E$-GOS) curves, would be used to discern the evolution from rotational to vibrational structure in nuclei as a function of spin. In addition, in order to get the insight into the rotational-like properties of nuclei, theoretical calculations have been performed for the yrast bands of the odd-A rare-earth nuclei using the total Routhian surfaces (TRS) model. The TRS plots indicate that the $^{165}$Yb and $^{157}$Dy isotopes have stable prolate shapes at low spin states. At higher rotational frequency ($\hbar\omega > 0.50\ MeV$), a distinct decrease in the quadrupole deformation is predicted by the calculations, and the isotopes becomes rigid in the $\gamma$ deformation.

## 1 Introduction

The phase transition with nucleon number and spin is one of the most significant topics in nuclear structure research. This transition is intimately related to the mechanisms by which atomic nuclei generate angular momentum. Recently, an abundance of observed phenomena connected with different collective band structures are well established by means of in-beam $\gamma$ –ray spectroscopy[1-3]. These structure characters manifest their angular momentum generation in different ways, and the different characteristics may even be involved in one mode of collective motion[4]. In nuclei where the excited state is generally formed by collective motion and nucleon pair breaking, a subtle rearrangement of only a few nucleons among the orbitals near the Fermi surface can result in completely different collective modes. Studying the microscopic excitations of many-nucleon systems and transitions between different excitation modes can shed light on the nature of the states lying close to the Fermi surface, allowing the shape of the nuclear mean field to be inferred. Regan et al.[5] proposed a simple method, named as the E-GOS (E-Gamma Over Spin) curve, for discerning the collective excitations resulting in oscillations or rotations. This prescription has been applied to the yrast cascades in the even-even nuclei and a clear structure evolution from vibration to rotation while spin increasing has been found[5]. Subsequently, the similar phenomena have also been discovered in other mass region[6]. In this work, however, we attempt to investigate whether the inverse process will occur or not.

*Corresponding author (email: zhb@mailbox.gxnu.edu.cn)

As is well known, nuclei in the mass region 150<$A$<190 are thought belonging to the well-deformed region, and then one may infer that the yrast band consequently exhibits a rotational structure. Recent findings[7], however, have given evidence for a particularly interesting phenomenon, i.e., the evolution from rotation to vibration as a function of spin, observed in the excitation spectra of even-even nuclei in this mass region. For an odd-$A$ nuclide, its high-spin states may be formed by coupling weakly the valence nucleon to the respective core excitations. Therefore, we may have a good chance to observe the similar phase evolution as their neighboring even-even nuclei in the region 150<$A$<190. In this paper, we aim at discussing the collective motions of an odd-$A$ nucleus for different spin ranges, and the mechanism of this phase transition will be discussed in the framework of the cranked shell model.

## 2 The E-GOS curve method

The concept of the E-GOS prescription has been applied to discern the structure evolution from vibration to rotation in nuclei as increasing spin[5]. In this method, the ratio of $E_\gamma(I \to I-2)/I$ can provide an effective way of distinguishing between axially symmetric rotational and harmonic vibrational modes[5]. For a vibrator, this ratio gradually diminishes to zero as the spin increases, while for an axially symmetric rotor it approaches a constant, $4[\hbar^2/2J]$. Here $J$ is the static moment of inertia. As an example, in Fig. 1(a) we show the E-GOS curves for a perfect harmonic vibrator and axially symmetric rotor with assuming the first excitations of 500 and 100 keV, respectively. This prescription can be used as a quite good signature to discern the vibrational to axially rotational structure evolution along the yrast line, and vice versa one can infer that this simple method can also manifest the transition from axially rotational to vibrational evolution. For the reader's convenience, the typical corresponding E-GOS curve for the axially rotational to vibrational structure evolution along the yrast line is also presented in Fig. 1(b), which is just the same as the Fig. 1(c) in Ref. [7]. In odd-$A$ systems, however, the effect of the bandhead spin should be taken into account. Then, the E-GOS prescriptions can be addressed by substituting the spin ($I$) by a normalized spin minus the bandhead spin projection on the axis of symmetry, $K$, such that $I \to (I-K)$. For good rotors, the E-GOS prescription for odd-$A$ systems then becomes[8]

$$R(I) = \frac{E_\gamma}{I} \to \frac{\hbar}{2J}\frac{(4I-2)}{I} \to \frac{\hbar^2}{2I}\frac{[4(I-K)]-2}{(I-K)}, \quad (1)$$

$$R(I-K) = \frac{E_\gamma - (4K\frac{\hbar^2}{2J})}{I-K} = \frac{E_\gamma - KR_{K+2}}{I-K}. \quad (2)$$

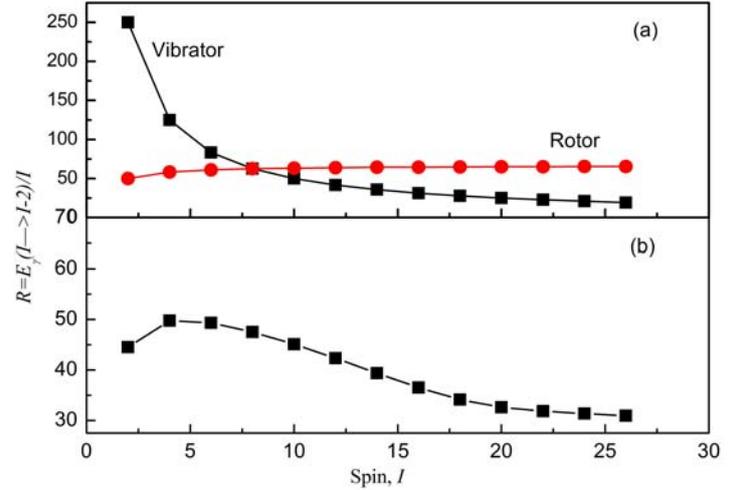

**Figure 1** (a) E-GOS curves for a perfect harmonic vibrator and axially symmetric rotor. (b) The characteristic of E-GOS plot for the axially rotational to vibrational shape transition with spin increasing.

Fig. 2 shows the E-GOS plots for the odd-$N$ isotopes in the mass region around $A$=160. These data are taken from refs. [9-29]. By comparing with the E-GOS curve characteristics presented in Fig.1(b), the E-GOS plots of $^{155,157,159,161}$Dy and $^{159,167}$Er shown in Fig. 2 present a clear evolution from rotational to vibrational excitations along the yrast line with increasing spin. Normally the $^{165,167}$Yb and $^{171}$Hf nuclei given in Fig.2 are thought belonging to the well-deformed region, and the good rotational energy spectra should be observed in these nuclei, and then every nucleus consequently exhibits a band structure with the yrast rotational band. So it is interesting to see that these yrast bands have the vibrational characteristic in the lower-spin region, whereas at higher spins it has a rotational pattern. Particularly, it should be mentioned that the results for the highest spins (last four points) of $^{167}$Yb, $^{171}$Hf and (last three points) of $^{165}$Yb in Fig. 2 point to an interesting aspect, the E-GOS curve in this region changes into the hyperbola expected for a vibrator. This suggests that above spin 35 there is again a change to the vibrational region in these three nuclei. In other nuclei, such as $^{157,159}$Gd and $^{167}$Hf, the vibrational characters were observed at low spins. Unfortunately, the experimental information on the high spin states for these bands isn't available now.

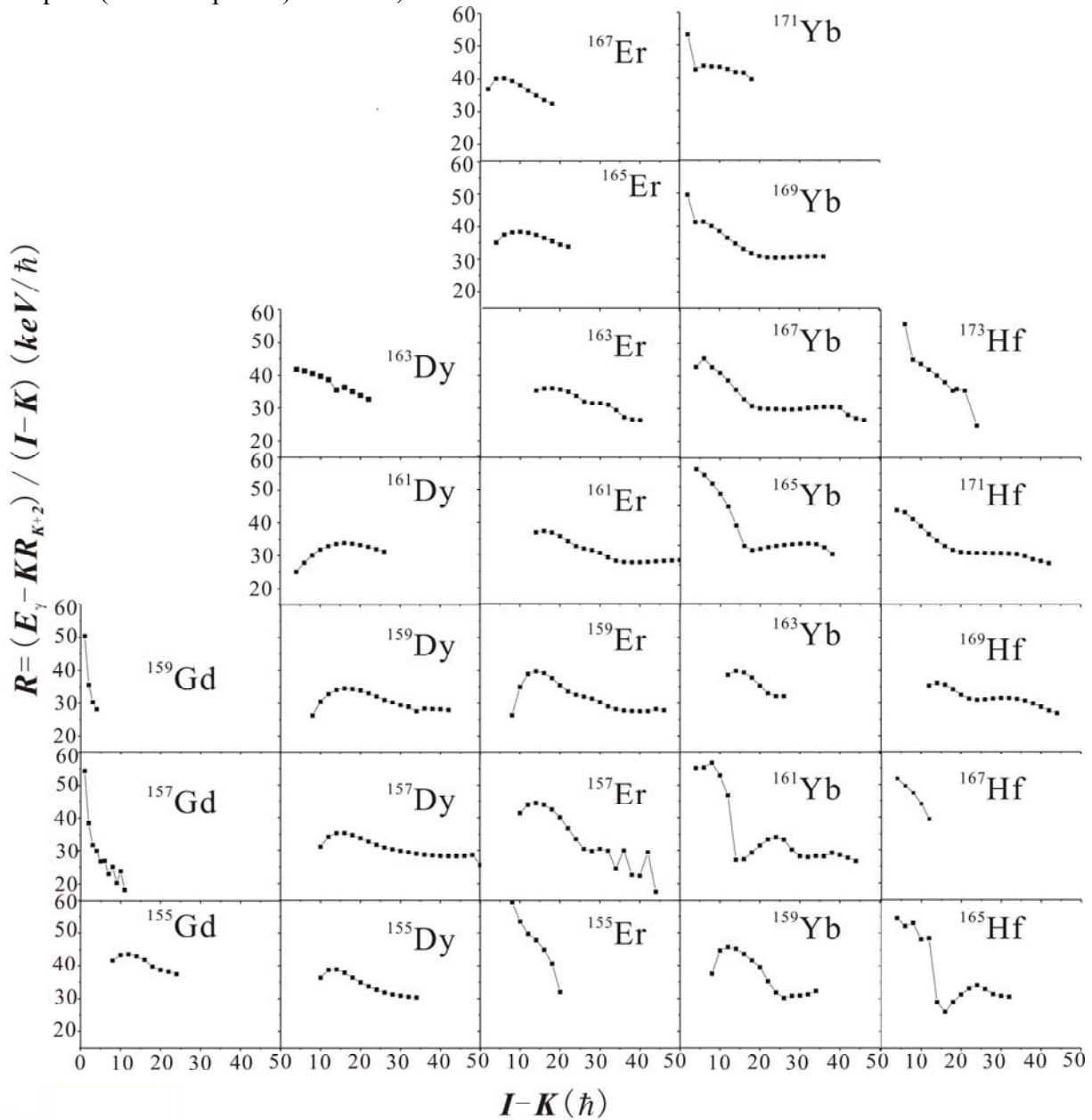

**Figure 2** The E-GOS curves of the odd-A nuclei in the mass region around A=160. The data are taken from Refs. [9–29].

## 3 Calculations and discussions

It was well known that the nuclear shapes are susceptible to the increased angular momentum. To gain an insight into the phase transitions and understand systematically the microscopic origin of these interesting phenomena, we have performed total-Routhian-surface (TRS) calculations based on the nonaxial deformed Woods-Saxon potential[30] in a three-dimensional deformation space ($\beta_2$, $\gamma$, $\beta_4$)[31]. In this method, both monopole and quadrupole pairings were included[32, 33]. At a given frequency, the deformation of a state is determined by minimizing the calculated TRS.

The shape calculations for two types of transition in the yrast bands of $^{157}$Dy and $^{165}$Yb were shown in Figs. 3 and 4, respectively. The calculations indicate that the low-lying configurations in $^{157}$Dy exhibit a stable prolate deformation but with a large triaxial shape ($\gamma \approx 22^0$). At higher rotational frequency ($\hbar\omega > 0.5$ MeV), a distinct decrease in the quadrupole deformation is predicted by the calculations. The same result was reported in Ref. [34] according to the measurement of B(E2) values. It was found that the quadrupole deformation ($\beta_2$) in the yrast band of $^{157}$Dy increases with increasing rotational frequency at low spin states. At higher spins ($\hbar\omega > 0.33$ MeV), however, the sudden reduction of $\beta_2$ values was also confirmed, which is consistent with the result presented in Fig. 3. As shown in Fig. 4, the nucleus $^{165}$Yb is predicted to be prolate with a quadrupole deformation of $\beta_2$=0.266 and a triaxiality parameter of $\gamma = -120^0$ at rotational frequency $\omega = 0$, which corresponds to ground state in $^{165}$Yb. In addition, it should be pointed out that when the rotational frequency $\omega = 0$, i.e., the nucleus is static, the $\gamma = -120^0$ is equivalent to $\gamma = 0$, namely axial symmetric with prolate shape. With spin increasing, a stable prolate deformation was indicated in $^{165}$Yb, which is in accordance with the rotational band structures observed in experiment[28]. However, a distinct decrease in the quadrupole deformation is predicted at $\hbar\omega > 0.60$ MeV by the calculations, and the shape of $^{165}$Yb develops to a triaxial prolate ($\gamma \sim -7.5^0$) deformation. This phenomenon can be ascribed to the rotation alignment of $i_{13/2}$ neutrons[28]. After band crossing, the effect of the quasi-particle excitations becomes important in excitation spectrum, which would result in reducing the collectivity of the nucleus[28]. This is consistent with the E-GOS curve property observed in $^{165}$Yb presented in Fig. 2.

As discussed above, the present TRS calculations indicate that the rotation vibration model can be used to describe the property of a stable deformed nucleus very well. In present work, however, it was found that the nuclear shape is strongly angular momentum dependent. Therefore, this case highlights the potential dangers of simply assuming the rotational model over the entire spin range. Moreover, the energies of various levels in the rare-earth nuclei are considered to follow very closely the simple formula[7]

$$E_I = \frac{\hbar^2}{2J}I(I+1) - BI^2(I+1)^2, \qquad (3)$$

where $J$ is an effective moment of inertia. Indeed, it is proved that the rotational collective sates can be approximated well enough, especially for the lower values of $I$, by the simpler formula

$$E_I = \frac{\hbar^2}{2J}I(I+1). \qquad (4)$$

This situation is assumed to be analogous to that of a diatomic molecule where rotational spectra with energies given by equation identical to Eq. (4) are known to exist[7]. But in our current work, a clear structure evolution with increasing spin is found in the odd-A nuclei by using the E-GOS prescription. So it's unreasonable to omit the second term of Eq. (3) for energy calculations if we consider all spin states.

The phase transition in rare-earth nuclei can be understood microscopically by the changes in the single-particle structure caused by the Coriolis force, which acts on the quasi-particles or nucleon pairs and leads to the alignment of the single-particle angular momenta along the rotation axis. As the influence of the Coriolis force is strongest on the high-$j$ with small $\Omega$ value orbits, the alignment of $i_{13/2}$ neutrons and $h_{11/2}$ protons are expected to be favored in this mass region. After band crossing, the rearrangement of the nuclear mass distribution and the polarizing effect from the quasi-particles would change the nuclear deformation and result in a loss of the axial symmetry of the nucleus. Particularly, the centrifugal force is also expected to influence the deformations in fast rotating nuclei.

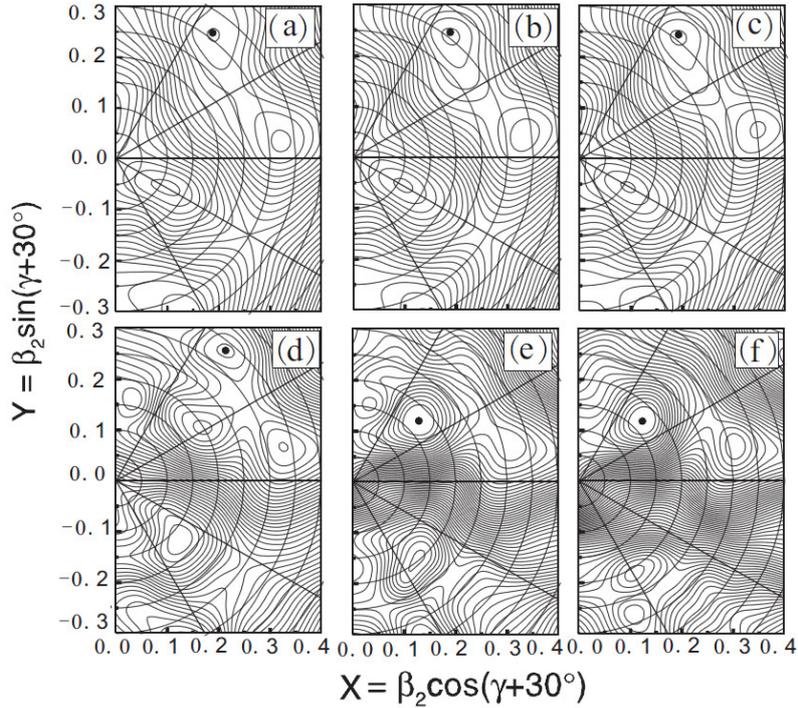

**Figure 3** Calculated total Routhian surfaces for the lowest $(\pi, \alpha) = (+, +1/2)$ configuration of $^{157}$Dy. The energy contours are at 200 keV intervals. The deformation parameters for the individual minima are: (a) $\hbar\omega= 0.0$ MeV, $\beta_2 = 0.311$, $\beta_4 = -0.090$, and $\gamma= 22.838°$ ; (b) $\hbar\omega= 0.10$ MeV, $\beta_2 = 0.309$, $\beta_4 = -0.091$, and $\gamma= 22.499°$ ; (c) $\hbar\omega= 0.20$ MeV, $\beta_2 = 0.310$, $\beta_4 = -0.090$, and $\gamma= 21.756°$ ; (d) $\hbar\omega= 0.40$ MeV, $\beta_2 = 0.340$, $\beta_4 = -0.089$, and $\gamma= 22.604°$ ; (e) $\hbar\omega= 0.60$ MeV, $\beta_2 = 0.176$, $\beta_4 = -0.097$, and $\gamma= 13.175°$; (d) $\hbar\omega= 0.80$ MeV, $\beta_2 = 0.171$, $\beta_4 = -0.097$, and $\gamma= 13.765°$.

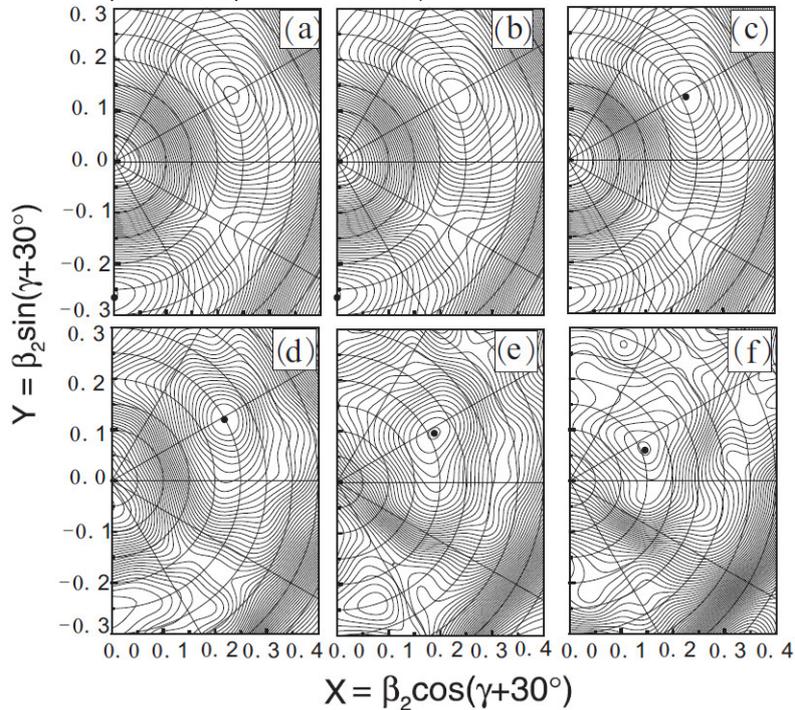

**Figure 4** Calculated total Routhian surfaces for the lowest $(\pi, \alpha) = (-, +1/2)$ configuration of $^{165}$Yb. The energy contours are at 200 keV intervals. The deformation parameters for the individual minima are: (a) $\hbar\omega= 0.0$ MeV, $\beta_2 = 0.266$, $\beta_4 = 0.011$, and $\gamma=-120°$ ; (b) $\hbar\omega= 0.10$ MeV, $\beta_2 = 0.266$, $\beta_4 = 0.012$, and $\gamma= -119.914°$ ; (c) $\hbar\omega= 0.20$ MeV, $\beta_2 = 0.260$, $\beta_4 = 0.007$, and $\gamma= -1.46°$ ; (d) $\hbar\omega= 0.40$ MeV, $\beta_2 = 0.249$, $\beta_4 = -0.012$, and $\gamma=-1.166°$ ; (e) $\hbar\omega= 0.60$ MeV, $\beta_2 = 0.211$, $\beta_4 =-0.022$, and $\gamma= -2.896°$ ; (f) $\hbar\omega= 0.80$ MeV, $\beta_2 = 0.158$, $\beta_4 = -0.032$, and $\gamma= -7.525°$ .

## 4 Conclusions

A simple prescription was used to distinguish vibrational from rotational regimes in the odd-A rare-earth nuclei. The characteristics of E-GOS curves of $^{155,157,159,161}$Dy and $^{159,167}$Er suggest that these nuclei undergo a clear evolution from rotational to vibrational excitations along the yrast line with increasing angular momentum. For the $^{165,167}$Yb and $^{171}$Hf nuclei, however, the E-GOS curves of these yrast bands have the vibrational characteristic in the lower-spin region, whereas at higher spins it has a rotational pattern, and above spin 35 there is again a change to the vibrational region. The total-Routhian-surface calculations with nonaxial deformed Woods-Saxon potential were performed to the analysis of shape evolution occurring in $^{157}$Dy and $^{165}$Yb. Comparison with the experimental data provides a consistent picture of the shape evolution in these nuclei in term of angular momentum. The current work also highlights the potential dangers of simply assuming the rotational-based concepts over the entire spin range. In this paper, we aim to discern the structure evolution in some mass region. The theoretical studies to describe such evolution from axial rotation to vibration with increasing spin in individual nucleus are beyond the scope of this work.

*This work was supported by the National Natural Science Foundation of China (Grants No. 11505035, 11465005) and Natural Science Foundation of Guangxi (Grant No. 2017GXNSFAA198160, 2015GXNSFDA139004).*